\documentclass[9pt,conference]{IEEEtran}

\usepackage{waspaa25} 

\usepackage{bm} 
\usepackage{cite}

\newcommand\linesubsec[1]{\vspace{0.8mm}\noindent\textbf{#1 --- }}

\title{Modulation Discovery with Differentiable Digital Signal Processing}

\name{Christopher Mitcheltree$^\flat$\thanks{Christopher Mitcheltree is supported jointly by UK Research and Innovation\\\indent(grant number EP/S022694/1) and Queen Mary University of London.},
      Hao Hao Tan$^\sharp$,
      Joshua D. Reiss$^\flat$}
\address{
    $^\flat$Centre for Digital Music, Queen Mary University of London, UK\\
    $^\sharp$Independent Researcher, Singapore
}

\begin{document}

\maketitle

\begin{abstract}
Modulations are a critical part of sound design and music production, enabling the creation of complex and evolving audio.
Modern synthesizers provide envelopes, low frequency oscillators (LFOs), and more parameter automation tools that allow users to modulate the output with ease.
However, determining the modulation signals used to create a sound is difficult, and existing sound-matching / parameter estimation systems are often uninterpretable black boxes or predict high-dimensional framewise parameter values without considering the shape, structure, and routing of the underlying modulation curves.
We propose a neural sound-matching approach that leverages modulation extraction, constrained control signal parameterizations, and differentiable digital signal processing (DDSP) to discover the modulations present in a sound.
We demonstrate the effectiveness of our approach on highly modulated synthetic and real audio samples, its applicability to different DDSP synth architectures, and investigate the trade-off it incurs between interpretability and sound-matching accuracy.
We make our code and audio samples available and provide the trained DDSP synths in a VST plugin.
\end{abstract}

\section{Introduction}
\label{sec:intro}


A key aspect of audio production is introducing dynamic variation and movement. 
This is achieved through \textit{modulation} (mod.), which uses one signal to control a parameter of another signal or component within a synthesizer, allowing the sound to evolve over time.
In modern synthesizers, this may involve an envelope generator modulating amplitude to shape a sound’s attack and decay, or a low-frequency oscillator (LFO) modulating parameters such as pitch (vibrato), filter cutoff (wah-wah), or amplitude (tremolo).
Modulation is ubiquitous in contemporary sound design and is especially crucial in electronic music, where artists continually push sonic boundaries.
For reference, $\sim 98\%$ of the presets in the default library of \texttt{Serum}\footnote{\texttt{https://xferrecords.com/products/serum-2}}, a popular commercial synthesizer, use modulation. 
Modern soft synths like \texttt{Vital}\footnote{\texttt{https://vital.audio/}} and \texttt{Serum} emphasize modulation as a core feature, offering drawable XY modulation grids and advanced modulation routing options such as wavetable position and spectral warping.

Parameter estimation for sound synthesizers, also known as \textit{sound matching}, is an active area of research focused on predicting synth parameters that best reproduce a target sound~\cite{heise2009automatic, tatar2016automatic, shier2021synthesizer, steinmetz2025audio}. 
Some approaches are broadly applicable across all types of synthesizers~\cite{yee-king2018automatic, esling2019flow, yang2023white, bruford2024synthesizer, hayes2025audio}, while others focus on specific techniques such as additive~\cite{engel2020ddsp, masuda2023improving}, wavetable~\cite{mitcheltree2021serumrnn, shan2022differentiable, ycy2024golf}, frequency modulation~\cite{caspe2022ddx7, chen2022sound2synth}, waveshaping~\cite{hayes2021neural}, and modular synthesis~\cite{turian2021one, uzrad2024diffmoog}.
Differentiable digital signal processing (DDSP) based methods formulate the parameter estimation problem as a self-supervised learning task, wherein a differentiable synthesizer is optimized to re-construct a target audio signal, and this process implicitly infers the underlying synth parameters. 
This class of methods shares similar objectives with the field of neural audio synthesis~\cite{hayes2024review}.

However, current research does not address the extraction and discovery of \textit{interpretable} modulation signals. 
Most works produce static, global synth parameters with no variation across time, or infer framewise mod. signals that are high-dimensional, difficult to interpret, or overly complex, which are in contrast to the intuitive modulation curves crafted by human sound designers. 
While parallel work exists on extracting time-varying control signals for audio effects \cite{mitcheltree2023modulation, ycy2024diffapf}, there has been limited exploration of this approach in the area of sound synthesis.
Motivated by this research gap, and inspired by related work on audio effect discovery~\cite{steinmetz2021steerable} and differentiable curve parameterizations~\cite{ha2018neural, nakano2019neural, scharwachter2021differentiable}, we propose a technique for discovering interpretable modulation signals present in a sound.

\begin{figure}[t]
  \centering
  \centerline{\includegraphics[trim=0.7cm 0cm 0.2cm 0cm, clip, width=\columnwidth]{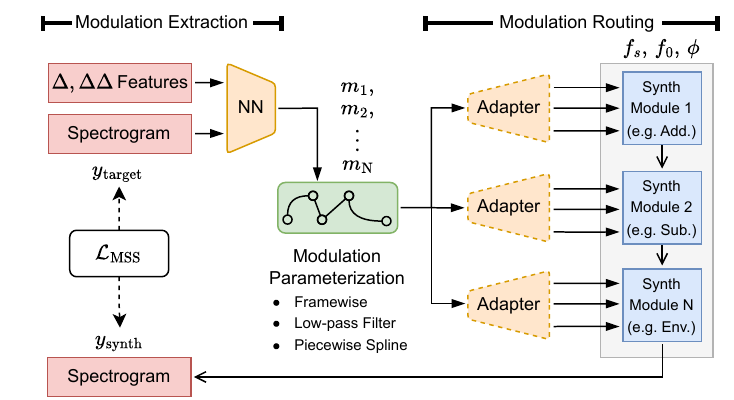}}
  \vspace{-5pt}
  \caption{Overview of the modulation discovery process through modulation extraction, parameterization, and routing using a DDSP synth. Orange blocks are neural networks, dashed blocks are optional, and blue blocks are differentiable and may contain learnable weights for sound matching.}
  \label{fig:modulation_discovery}
\end{figure}

\noindent Our contributions are as follows:
\begin{itemize}
    \item We introduce a modulation discovery approach (see Figure~\ref{fig:modulation_discovery}) that leverages modulation extraction, DDSP synthesis, and self-supervised sound matching to implicitly infer underlying modulations in arbitrary target audio.
    \item We define constrained, differentiable modulation signal parameterizations based on low-pass filtering and splines that ensure interpretable curves resembling those made by sound designers.
    \item We demonstrate the effectiveness of our approach on reconstructing highly modulated audio samples, both synthetic and real, and investigate the trade-off between mod. signal interpretability and sound matching accuracy for different DDSP synth architectures.
\end{itemize}

\begin{figure}[t]
  \centering
  \centerline{\includegraphics[trim=3.8cm 0.2cm 0cm 1.7cm, clip, width=0.93\columnwidth]{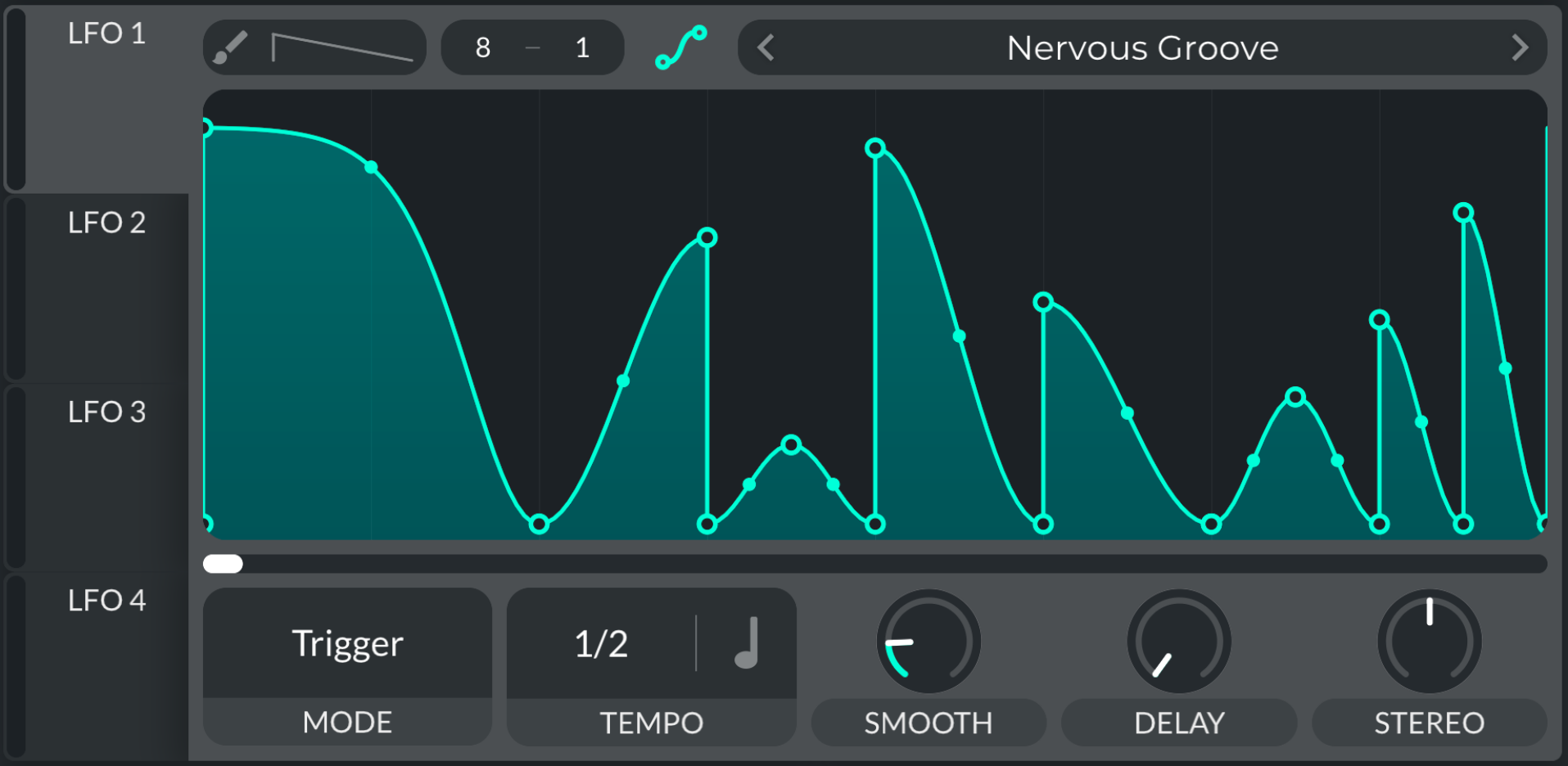}}
  \vspace{-5pt}
  \caption{A 2D drawable modulation grid in the \texttt{Vital} soft synth.}
  \label{fig:vital_mod_grid}
  \vspace{-5pt}
\end{figure}

\clearpage

\section{Methodology}
\label{sec:methodology}

We propose a modulation discovery approach (shown in Figure~\ref{fig:modulation_discovery}) consisting of three steps: modulation routing, extraction, and parameterization, that allows interpretable mod. signals to be uncovered in black / gray-box time-varying audio sources.
While this work focuses on modulations of software synths for electronic music, our approach is general and can be applied to other instruments and audio domains.

\subsection{Modulation Routing with DDSP}
\label{ssec:mod_routing}

First, a DDSP synth architecture is selected that consists of controllable, time-varying synth modules and reflects the routing and high-level types of modulations to be discovered.
In this work, we focus on additive (Add.), subtractive (Sub.), and envelope (Env.) modulations. 
Leveraging DDSP provides interpretability, computational efficiency, and strong guardrails on generated audio. 
Inspired by the popularity and expressiveness of soft synths like \texttt{Serum} and \texttt{Vital}, we build a synth (called Mod. Synth) with three differentiable modules in series: a wavetable oscillator with controllable wavetable position, a resonant filter with controllable coefficients, and an envelope. 
We implement the wavetable module using \texttt{PyTorch}'s \texttt{grid\_sample} function and use the time-varying biquad filter from~\cite{ycy2024diffapf} for the filter module.

\subsection{Modulation Extraction through Sound Matching}
\label{ssec:mod_extraction}

Next, a neural network extracts low-dimensional (in this work 1D) mod. signals for each synth module via a sound matching setup.
We use LFO-net from~\cite{mitcheltree2023modulation} which has a high inductive bias for extracting modulations and can be trained with limited data.
A simple multilayer perceptron adapter converts these mod. signals to higher dimensions for synth modules that require it (e.g. filter coefficients).
Training is self-supervised where input and output audio is compared using a perceptual loss function.
This information bottleneck + adapter technique lowers sound matching accuracy, but provides interpretability and causal controllability of the synth during inference.
Leveraging sound matching to discover modulations means this approach can be applied to any evolving audio and has the flexibility to uncover modulation shapes and combinations different from those present in the target audio.


\subsection{Modulation Parameterization}
\label{ssec:mod_parameterization}

\begin{figure}[t]
  \vspace{-5pt}
  \centering
  \centerline{\includegraphics[width=\columnwidth]{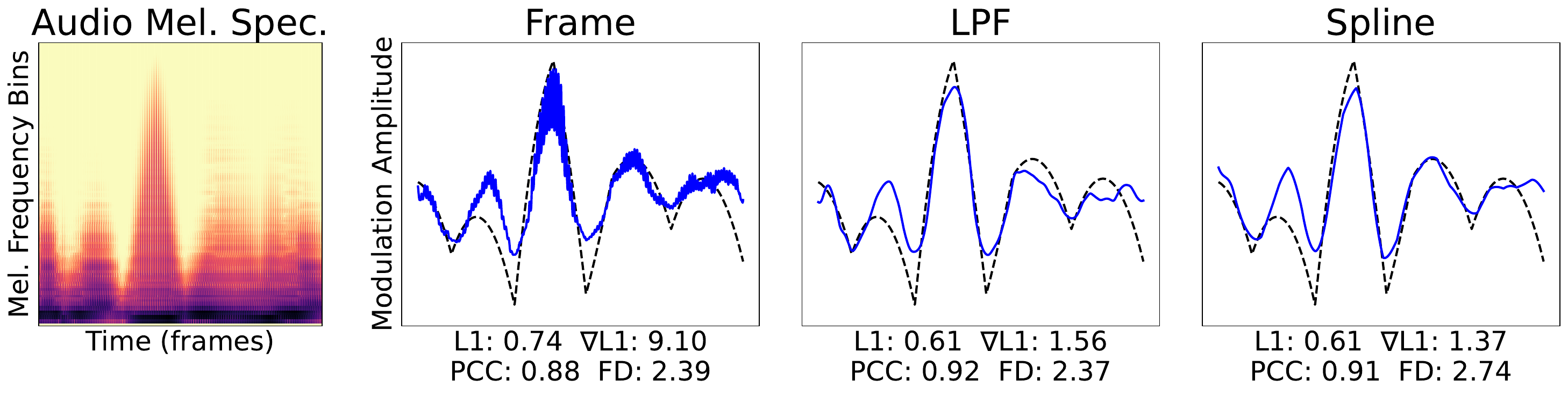}}
  \vspace{-8pt}
  \caption{Modulation extraction with the framewise, low-pass filtered, and piecewise Bézier curve parameterizations. The ground truth signal is dashed and the four mod. signal distance measures from Section~\ref{ssec:modulation_extraction} are computed.}
  \vspace{-5pt}
  \label{fig:parameterizations}
\end{figure}

Finally, we propose three differentiable parameterizations for the extracted low-dimensional mod. signals and investigate their curve fitting and sound matching abilities in the experiments section.
Figure~\ref{fig:parameterizations} provides a visualization of them and modulation extraction.

\linesubsec{Framewise (Frame)}
By default, mod. signals are extracted frame-by-frame at a control rate $f_{\text{MS}}$ much slower than the sample rate\\(i.e. $f_{\text{MS}} \approx \frac{f_s}{100}$).  
Most DDSP systems use this parameterization.

\linesubsec{Low-pass Filter (LPF)}
We post-process extracted mod. signals at $f_{\text{MS}}$ with a low-pass filter and cutoff frequency $f_c < 20$~Hz, which ensures they are smooth and similar to LFO signals.

\linesubsec{2D Bézier Curve (Spline)}
Inspired by the drawable XY modulation grids of soft synths (shown in Figure~\ref{fig:vital_mod_grid}), we implement a piecewise 2D Bézier curve~\cite{bezier1968procede} mod. signal parameterization $B(t)$.

\noindent We define $K \in \mathbb{Z}_{>0}$ segments on intervals $u_k$ with each segment $B_{k}$ consisting of $n \in \mathbb{Z}_{>0}$ control points $P_{k,i}$ such that:
\begin{align}
    P_{k,i} = (x_{k,i}, y_{k,i}) \in \mathbb{R}^2, \quad i = 0, 1, \dots, n \\
    B_{k}(u_{k}) = \sum_{i=0}^{n} \binom{n}{i}(1 - u_{k})^{n-i} u_{k}^i\,P_{k,i} \\
    \text{where} \enspace k = \text{min}(\lfloor K \cdot t \rfloor, K - 1), \quad u_{k} = K \cdot t - k \\
    B(t) = B_{k}(u_{k}), \quad t \in [0,1]
\end{align}

\noindent To ensure $C^0$ (positional) continuity and covering all of $t$, we enforce
\begin{align}
P_{k,n} = P_{k+1,0}, \quad \forall k = 0,1,\dots,K-2 \\
0 = x_{0,0} < x_{0,1} < \cdots < x_{0,n} = x_{1,0} < \cdots < x_{K-1,n} = 1
\end{align}

\noindent Bézier curves offer several advantages, including compact, low-dimensional representations of smooth signals, differentiability and efficient computation due to their closed-form expression, and interpretability with bounded outputs, as the curve remains entirely within the convex hull of its control points.



\section{Experiments}
\label{sec:experiments}

We conduct three experiments to evaluate our proposed approach: its \emph{modulation extraction} capabilities on unseen real-world mod. signals (Sec.~\ref{ssec:modulation_extraction}), and its \emph{modulation discovery} ability for synthetic (Sec.~\ref{ssec:modulation_discovery_synthetic}) and real-world audio (Sec.~\ref{ssec:modulation_discovery_real}).
We provide audio samples, VST plugins, code, and additional details at the accompanying website.\footnote{\texttt{https://christhetree.github.io/mod\_discovery/}}

\linesubsec{Training}
All experiments train LFO-nets with 1.0$\pm$0.2M parameters and a temporal receptive field of 1471 frames for 30 epochs on a 60-20-20\% train-val-test split, with a batch size of 32, 
the Schedule-Free AdamW optimizer~\cite{defazio2024road}, a learning rate of 0.005 for models / 0.05 for learnable synth modules, and using multi-resolution STFT loss (MSS)~\cite{yamamoto2020parallel}.
Spline mod. signals are linearly interpolated over the first $\frac{2}{3}$ of training from linear to cubic which encourages learning high-level structure before becoming detailed and curvy.
In Experiments~\ref{ssec:modulation_discovery_synthetic} and \ref{ssec:modulation_discovery_real} for all parameterizations, Gaussian noise with $\sigma = 0.33$ is added (and linearly decayed over the first $\frac{2}{3}$ of training) to the predicted wavetable position of Mod. Synth to encourage the use of all available learnable wavetable positions.

\linesubsec{Input Features} Input to all models is a 3-channel Mel spectrogram (128 Mel bins, 1024 FFT size, 96 sample hop length) with 50~ms delta and delta-delta features~\cite{furui1986speaker} computed from three seconds of $f_s = 48$~kHz audio which results in 1501 frames and a $f_{\text{MS}}$ of 500~Hz.
Since pitch modulations are not being investigated, the $f_0$ of the input audio is provided to synths and phase ($\phi$) is randomized.

\subsection{Modulation Extraction (Real-world Data)}
\label{ssec:modulation_extraction}

The first experiment evaluates how well LFO-net and the three parameterization methods can extract mod. signals when a white-box synth architecture is used.
Audio is generated and reconstructed with a frozen Mod. Synth (see Section~\ref{ssec:mod_routing}), which means extracted mod. signals can be compared directly to their ground truth counterparts.

\linesubsec{Parameterization}
We use a cubic 2D Bézier curve with 24 segments for the spline method, so it can approximate up to an $\sim 8$~Hz sinusoidal mod. signal. 
For a fair comparison, we use a Blackman windowed-sinc low-pass filter with an $f_c = 8$~Hz cutoff frequency and $\left\lceil \frac{f_{\text{MS}}}{8} \right\rceil$ taps for the LPF method.

\linesubsec{DDSP Synth}
The Mod. Synth wavetable uses a LPF for anti-aliasing with $f_c = 0.9 \cdot \frac{f_s}{2}$.
The filter module is a low-pass biquad filter with $100~\text{Hz} \leq f_c \leq 8~\text{kHz}$ and $\frac{\sqrt{2}}{2} \leq Q \leq 4$.

\linesubsec{Datasets}
Models are trained on 1024 datapoints of three randomly generated spline mod. signals (one for each synth module) using a 1-3 degree piecewise 1D Bézier curve consisting of 1-8 segments.
Segment intervals are random with a min. length of half a uniform segment length.
Filter resonance values ($Q$) are constant and sampled log-uniformly.
10 wavetables from \texttt{Ableton Live 12}'s stock library are used for initializing Mod. Synth's additive module.
They consist of 4 to 256 positions, 1024 samples, and are chosen to be tonal.
10 models are trained for each wavetable using different random seeds. 

All 100 models are tested on an unseen, real-world mod. signal dataset consisting of modulation curves from \texttt{Vital}'s default preset library.
Signals that consist of $>24$ segments, span $<50\%$ of the modulation range, or are $>50\%$ flat are removed, leaving 52 unique curves.
1024 randomly sampled datapoints from the resulting $\sim 132$k unique combinations of three mod. signals are used for evaluation.

\linesubsec{Evaluation}
We compare Add, Sub, and Env. mod. signals using four different measures: $L_1$ distance, first-order central difference $L_1$ distance ($\nabla L_1$), Pearson correlation coefficient (PCC), and Fréchet distance (FD).
They are chosen to cover shift, scale, and time-warping invariance to compare shape, structure, absolute, and relative differences between two modulations.
We also calculate normalized total variation (TV), turning points (TP), and spectral entropy (SE) for each signal.
Uniformly sampled random spline and framewise baselines are included as an anchor to compare against.

\linesubsec{Results}
From Tables~\ref{tab:exp1_mod_distances} and \ref{tab:exp1_mod_metrics}, we observe that all parameterizations minimize absolute differences ($L_1$) between mod. signals. Unsurprisingly, Spline and LPF better capture smooth variations ($\nabla L_1$), with Spline best matching the target TV and TP metrics. LFO-net also excels at extracting envelope and filter modulations over wavetable, likely due to their simpler, more visible effect on the input features.

\begin{table}[t]
\centering
\caption{Modulation extraction evaluation results on unseen real-world data.}
\label{tab:exp1_mod_distances}
\sisetup{
    reset-text-series = false, 
    text-series-to-math = true, 
    mode=text,
    tight-spacing=true,
    table-format=1.2,
    table-number-alignment=center,
    separate-uncertainty = true,
    detect-weight = true,
    detect-inline-weight = math,
}
\setlength{\tabcolsep}{4.3pt}
\begin{tabular}{
    ll
    S[table-format=1.2(1.2)]
    S[table-format=3.2(1.2)]
    S[table-format=1.2(1.2)]
    S[table-format=1.2(1.2)]
}
    \toprule
    Mod. & Method & {$L_1 \downarrow$} & {$\nabla L_1 \downarrow$} & {PCC $\uparrow$} & {FD $\downarrow$} \\
    \midrule
    Add. & Frame  & \bfseries 2.41(0.08) &           2.51(0.06) & \bfseries 0.46(0.03) &           5.36(0.09) \\
         & LPF    &           2.42(0.07) &           1.78(0.03) & \bfseries 0.46(0.02) & \bfseries 5.29(0.07) \\
         & Spline &           2.44(0.09) & \bfseries 1.71(0.03) &           0.45(0.03) &           5.33(0.09) \\
    \midrule
    Sub. & Frame  &           1.45(0.01) &           2.04(0.03) & \bfseries 0.76(0.01) &           4.33(0.05) \\
         & LPF    & \bfseries 1.44(0.01) &           1.52(0.02) & \bfseries 0.76(0.01) &           4.28(0.03) \\
         & Spline &           1.45(0.01) & \bfseries 1.48(0.02) &           0.75(0.01) & \bfseries 4.27(0.04) \\
    \midrule
    Env. & Frame  & \bfseries 2.11(0.02) &           3.51(0.09) & \bfseries 0.68(0.01) &           5.25(0.06) \\
         & LPF    &           2.12(0.02) &           1.85(0.02) & \bfseries 0.68(0.01) &           5.22(0.06) \\
         & Spline & \bfseries 2.11(0.01) & \bfseries 1.76(0.02) & \bfseries 0.68(0.01) & \bfseries 5.18(0.06) \\
    \midrule
    All  & Rand. Spline  & 3.35(0.01) & 8.96(0.04)   & 0.00(0.01) & 7.10(0.05) \\
         & Rand. Frame   & 3.69(0.01) & 166.80(0.13) & 0.00(0.00) & 7.77(0.05) \\
    \bottomrule
\end{tabular}
\vspace{-4pt}
\end{table}

\begin{table}[t]
\centering
\caption{Extracted, target, and random baseline modulation signal metrics. Values closer 
 to their corresponding target values are better.}
\label{tab:exp1_mod_metrics}
\sisetup{
    reset-text-series = false, 
    text-series-to-math = true, 
    mode=text,
    tight-spacing=true,
    table-format=1.2,
    table-number-alignment=center,
    separate-uncertainty = true,
    detect-weight = true,
    detect-inline-weight = math,
}
\begin{tabular}{
    l
    S[table-format=3.2(1.2)]
    S[table-format=3.2(1.2)]
    S[table-format=1.3(1.3)]
}
    \toprule
    Method & {Total Variation} & {Turning Points} & {Spectral Entropy} \\
    \midrule
    Frame        &           4.40(0.09) &          157.01(2.79) &           0.490(0.002) \\
    LPF          &           2.28(0.01) &           14.46(0.20) & \bfseries 0.453(0.002) \\
    Spline       & \bfseries 2.10(0.02) & \bfseries 10.05(0.12) &           0.451(0.002) \\
    \midrule
    Target       &           1.69(0.01) &            1.21(0.06) &           0.464(0.006) \\
    Rand. Spline &           9.50(0.03) &           26.88(0.08) &           0.560(0.001) \\
    Rand. Frame  &         333.61(0.25) &          666.79(0.36) &           0.935(0.000) \\
    \bottomrule
\end{tabular}
\vspace{-4pt}
\end{table}

\subsection{Modulation Discovery (Synthetic Data)}
\label{ssec:modulation_discovery_synthetic}

The second experiment evaluates how well our modulation routing and DDSP sound matching approach can discover modulations for a gray-box synth.
This experiment is identical to Section~\ref{ssec:modulation_extraction}, except only synthetic mod. signals are used and now Mod. Synth has a learnable, 16-position, 1024-sample wavetable initialized with Gaussian noise ($\sigma = 0.01$) and the learnable time-varying biquad filter from~\cite{ycy2024diffapf}.

\linesubsec{Evaluation}
Discovered mod. signals for learnable synth modules can no longer be directly compared to their respective ground truth signals since they may be shifted, flipped, and scaled depending on the stochastic gradient descent optimization process.
For example, if the learned wavetable module happens to be in reverse order of the original in the dataset, then the discovered additive modulations will be mirrored vertically.
Therefore, we post-process extracted mod. signals with a linear least squares (LLS) solver using their corresponding ground truth signal as the target (labeled LLS 1).
Discovered mod. signals may also combine together and fulfill different purposes to match the target sound.
For example, some of the envelope modulations in the target audio may be achieved using a mixture of wavetable and filter modulations in the learned synth.
Consequently, we use the LLS solver to make a linear combination of all three discovered mod. signals for each ground truth signal (labeled LLS 3).
We then apply the same mod. signal distance metrics as before.

For measuring sound matching quality, we use MSS, Mel-frequency cepstral coefficients (MFCC) $L_1$ distance, and two Fréchet Audio Distances (FAD)~\cite{elizalde2023clap, defossez2023high} which have been shown to correlate with human perception~\cite{kilgour2019interspeech}.
We include the random spline baseline from before (mod. signals are only randomized during testing, not training) as well as an Oracle baseline that always uses the ground truth mod. signals and serves as a consistent point of comparison for evaluating the sound matching ability of the LFO-net + Mod. Synth system.

\linesubsec{Results}
Table~\ref{tab:exp2_mod_distances} shows that LLS 3 significantly outperforms LLS 1 across all signals and parameterizations, suggesting that the modulation discovery process distributes modulations across different synth modules. 
This is particularly evident for Sub. and Env., which make similar perceptual changes to the output audio. 
LPF and Spline again best match the ground-truth signals, benefiting from strong inductive biases. 
Compared to the previous experiment, extracted modulation evaluation distances are more consistent and uniformly distributed, highlighting the advantage of a flexible sound matching approach with learnable synth modules.
Looking at the audio similarity metrics in Table~\ref{tab:exp2_sound_matching}, all methods sometimes surpass the Oracle baseline on MSS and FAD. 
This indicates that discovered modulations can better optimize the perceptual MSS loss and that MFCC is the most reliable metric for evaluating underlying modulation structure in audio.

\begin{table}[t]
\centering
\caption{Modulation discovery evaluation results for synthetic data.}
\label{tab:exp2_mod_distances}
\sisetup{
    reset-text-series = false, 
    text-series-to-math = true, 
    mode=text,
    tight-spacing=true,
    table-format=1.2,
    table-number-alignment=center,
    separate-uncertainty = true,
    detect-weight = true,
    detect-inline-weight = math,
}
\setlength{\tabcolsep}{3.9pt}
\begin{tabular}{lll*{4}{S[table-format=1.2(1.2)]}}
    \toprule
    Mod. & Proc. & Method & {$L_1 \downarrow$} & {$\nabla L_1 \downarrow$} & {PCC $\uparrow$} & {FD $\downarrow$} \\
    \midrule
    Add. & LLS 1  & Frame  &           1.11(0.02) &           2.16(0.12) &           0.47(0.03) &           3.36(0.04) \\
         &        & LPF    & \bfseries 1.08(0.03) &           1.42(0.03) & \bfseries 0.50(0.02) & \bfseries 3.19(0.06) \\
         &        & Spline &           1.16(0.01) & \bfseries 1.37(0.03) &           0.43(0.02) &           3.27(0.03) \\
    \cmidrule(lr){2-7}
         & LLS 3  & Frame  &           0.91(0.02) &           3.08(0.13) &           0.67(0.01) &           3.38(0.13) \\
         &        & LPF    & \bfseries 0.88(0.02) &           1.58(0.04) & \bfseries 0.69(0.01) &           2.90(0.08) \\
         &        & Spline &           0.94(0.01) & \bfseries 1.53(0.03) &           0.66(0.01) & \bfseries 2.87(0.03) \\
    \midrule
    Sub. & LLS 1  & Frame  &           0.62(0.03) &           2.90(0.25) & \bfseries 0.82(0.01) &           2.84(0.18) \\
         &        & LPF    & \bfseries 0.61(0.03) & \bfseries 1.03(0.03) & \bfseries 0.82(0.02) &           2.27(0.12) \\
         &        & Spline &           0.67(0.04) &           1.04(0.06) &           0.79(0.03) & \bfseries 2.24(0.12) \\
    \cmidrule(lr){2-7}
         & LLS 3  & Frame  &           0.50(0.01) &           2.66(0.14) &           0.89(0.01) &           2.69(0.17) \\
         &        & LPF    & \bfseries 0.48(0.01) &           1.06(0.02) & \bfseries 0.90(0.01) &           1.97(0.05) \\
         &        & Spline &           0.52(0.01) & \bfseries 1.04(0.02) &           0.88(0.01) & \bfseries 1.88(0.03) \\
    \midrule
    Env. & LLS 1  & Frame  &           1.28(0.02) &           2.16(0.07) &           0.35(0.01) &           3.90(0.13) \\
         &        & LPF    &           1.27(0.02) &           1.43(0.04) &           0.37(0.01) &           3.61(0.07) \\
         &        & Spline & \bfseries 1.25(0.02) & \bfseries 1.39(0.03) & \bfseries 0.38(0.01) & \bfseries 3.54(0.06) \\
    \cmidrule(lr){2-7}
         & LLS 3  & Frame  &           0.96(0.02) &           3.74(0.14) &           0.67(0.01) &           3.62(0.18) \\
         &        & LPF    & \bfseries 0.92(0.02) &           1.68(0.04) & \bfseries 0.69(0.01) &           3.00(0.09) \\
         &        & Spline &           0.95(0.02) & \bfseries 1.60(0.03) &           0.68(0.01) & \bfseries 2.91(0.06) \\
    \midrule
    All     & LLS 1  & Rand. S.  & 1.38(0.02) & 1.91(0.04) & 0.15(0.01) & 3.65(0.04) \\
    \cmidrule(lr){2-7}
         & LLS 3  & Rand. S.  & 1.33(0.02) & 2.83(0.04) & 0.30(0.01) & 3.47(0.05) \\
    \bottomrule
\end{tabular}
\vspace{-15pt}
\end{table}

\begin{table}[t]
\vspace{-11pt}
\centering
\caption{Sound matching audio similarity metrics for synthetic data.}
\label{tab:exp2_sound_matching}
\sisetup{
    reset-text-series = false, 
    text-series-to-math = true, 
    mode=text,
    tight-spacing=true,
    table-format=1.2,
    table-number-alignment=center,
    separate-uncertainty = true,
    detect-weight = true,
    detect-inline-weight = math,
}
\setlength{\tabcolsep}{4.0pt}
\begin{tabular}{
    l
    S[table-format=2.2(1.2)]
    S[table-format=1.2(1.2)]
    S[table-format=3.2(2.2)]
    S[table-format=3.2(2.2)]
}
    \toprule
    Method & {MSS $\downarrow$}   & {MFCC $\downarrow$}  & {FAD MS-CLAP $\downarrow$} & {FAD EnCodec $\downarrow$} \\
    \midrule
    Frame    & \bfseries 0.94(0.01) &           1.60(0.02) & \bfseries 109.83(4.49) &           12.92(1.78) \\
    LPF      &           0.95(0.02) & \bfseries 1.54(0.03) &           115.70(7.13) & \bfseries 12.73(1.58) \\
    Spline   &           1.02(0.02) &           1.76(0.03) &           131.93(6.26) &           16.27(1.77) \\
    \midrule
    Oracle   &           1.05(0.03) &           1.48(0.06) &            111.43(8.83) &           16.80(1.12) \\
    Rand. S. &          10.51(6.25) &           4.67(0.09) &           764.73(55.00) &         181.60(85.49) \\
    \bottomrule
\end{tabular}
\vspace{-8pt}
\end{table}


\subsection{Modulation Discovery (Real-world Audio)}
\label{ssec:modulation_discovery_real}

The last experiment evaluates how well our modulation discovery approach generalizes to real-world audio, black-box synths, and different DDSP synth architectures and their modulation routings.

\linesubsec{Dataset}
This experiment is identical to Section~\ref{ssec:modulation_discovery_synthetic}, except the dataset now consists of complex, highly modulated audio from the popular soft synth \texttt{Serum}.
All preset patches in the ``Bass (Hard)'' category with both wavetable and filter modulations are selected and notes A0 to C8 are rendered, resulting in 44 minutes of audio and 880 datapoints total.
Three different DDSP synths and their models are trained 20 times on the dataset with unique random seeds.

\begin{figure}[t]
    \centering
    \begin{minipage}{1.0\linewidth}
        \centering
        \includegraphics[width=1.0\linewidth]{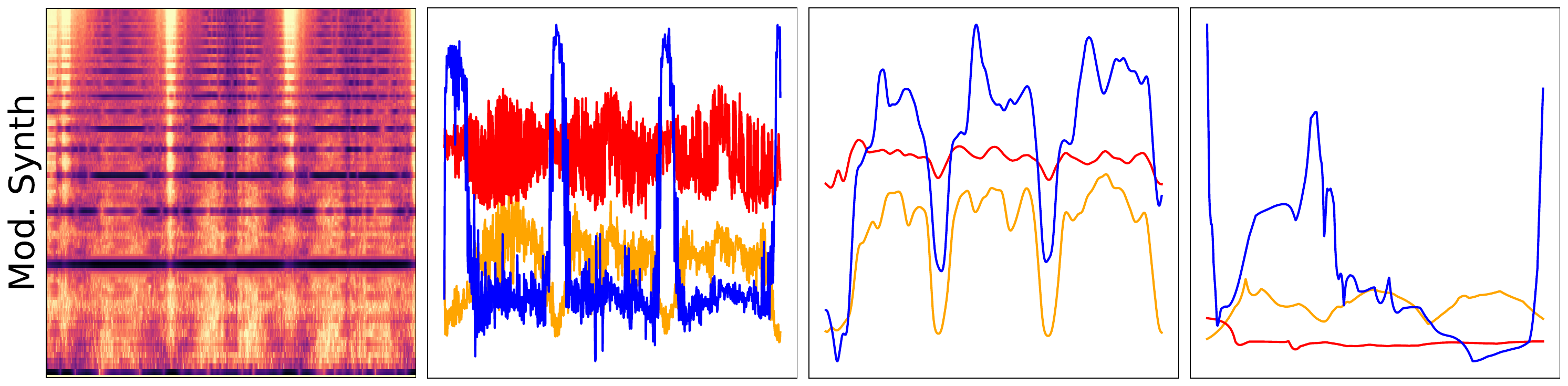}
    \end{minipage}
    \begin{minipage}{1.0\linewidth}
        \centering
        \includegraphics[width=1.0\linewidth]{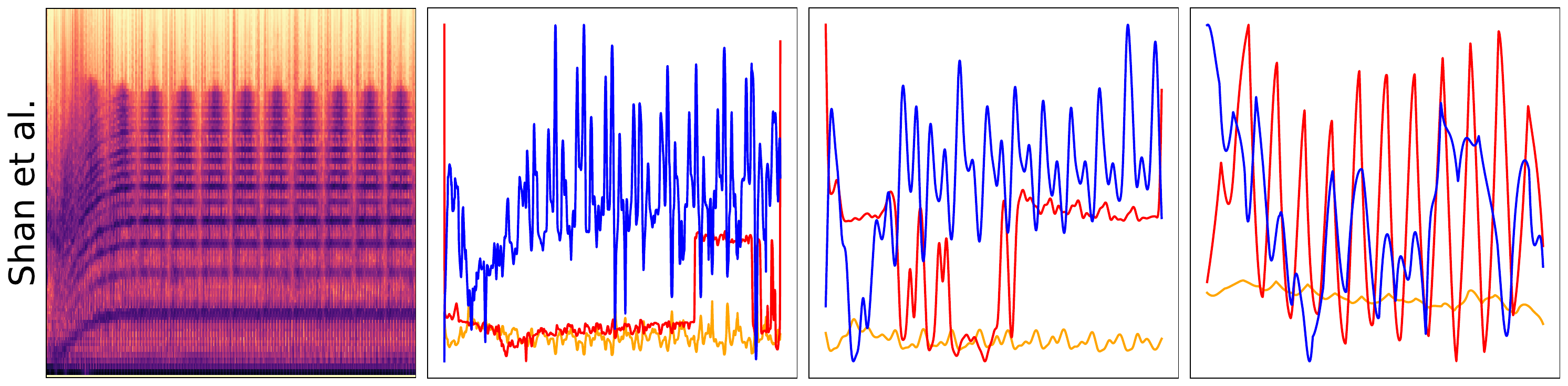}
    \end{minipage}
    \begin{minipage}{1.0\linewidth}
        \centering
        \includegraphics[width=1.0\linewidth]{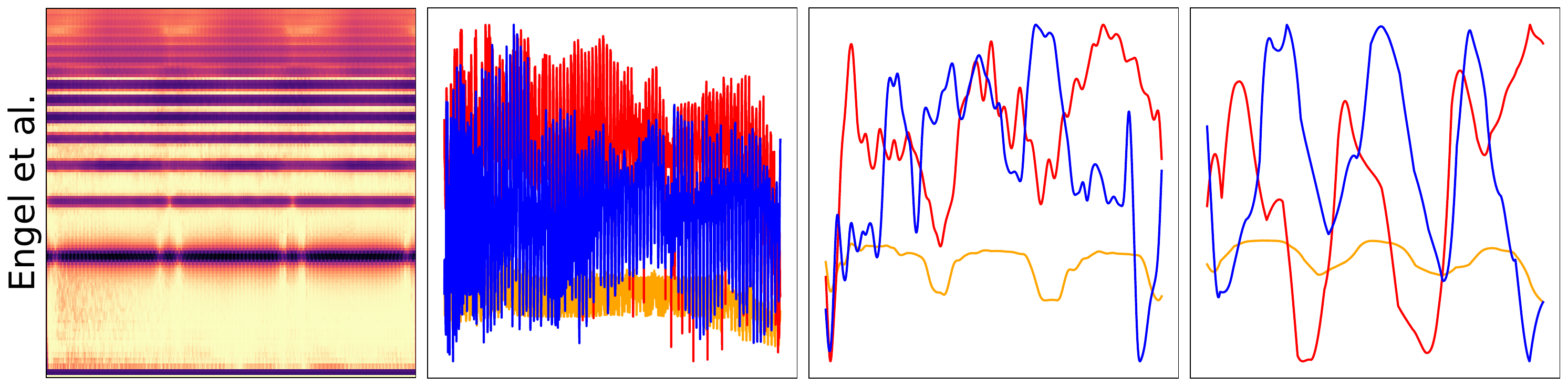}
    \end{minipage}
    \caption{Discovered additive (red), subtractive (blue), and envelope (orange) modulations in \texttt{Serum} test dataset audio samples using different DDSP synths and Frame (left), LPF (center), and Spline (right) mod. signal parameterizations.}
    \label{fig:serum_discovered_modulations}
\end{figure}

\linesubsec{Evaluation}
We test Mod. Synth, a variant of Mod. Synth with Shan et al.'s~\cite{shan2022differentiable} wavetable module which uses time-varying attention across wavetable positions, and Engel et al.'s~\cite{engel2020ddsp} harmonic oscillator + filtered noise synth (100 harmonics, 65 noise bands) with a learnable envelope at its output.
Since the underlying DSP mod. signals of the target audio are now unknown, we compute 1D audio features (root mean square loudness (RMS) and spectral flatness (SF)) at a frame rate of $f_{\text{MS}}$ to use as approximations and apply the LLS 3 post-processing step to extracted mod. signals before comparing.
We hypothesize that RMS should correlate with sub. and env. modulations, while SF is perceived as tonality and should correlate with the add. and sub. modulations.
To determine the trade-off between sound matching quality and interpretability incurred by extracting low-dimensional mod. signals for each synth module (see Section~\ref{ssec:mod_extraction}), we include a granular mod. signal baseline (Gran.) that uses a high-dimensional framewise mod. signal for its synth modules. 
This eliminates the information bottleneck and is the default setup in most DDSP systems.
Figure~\ref{fig:serum_discovered_modulations} visualizes some of the discovered modulations.

\linesubsec{Results}
Overall, sound matching and modulation extraction performance declines, reflecting the complexity of the \texttt{Serum} audio, the limited expressiveness of the selected synths, and the approximation of the ``ground-truth'' signals. 
Table~\ref{tab:exp3_mod_distances} shows LPF consistently yields the closest match to RMS and SF modulations across architectures, with Spline usually a close second.
Additional experiments with spectral centroid and bandwidth also give similar results.
Figure~\ref{fig:serum_discovered_modulations} illustrates how the Frame method overfits each timestep to maximize sound matching, a trend echoed in Table~\ref{tab:exp3_sound_matching}: parameterizations with more degrees of freedom better optimize perceptual loss, but sacrifice interpretability. 
Importantly, the drop in matching accuracy from Gran. to LPF is far smaller for Mod. Synth and Shan et al. than for Engel et al., underscoring the advantage of choosing a DDSP architecture with lower-dimensional control parameters for its modules.

\begin{table}[t]
\centering
\caption{Modulation discovery evaluation results for real-world data.}
\label{tab:exp3_mod_distances}
\sisetup{
    reset-text-series = false, 
    text-series-to-math = true, 
    mode=text,
    tight-spacing=true,
    table-format=1.2,
    table-number-alignment=center,
    separate-uncertainty = true,
    detect-weight = true,
    detect-inline-weight = math,
}
\setlength{\tabcolsep}{3.8pt}
\begin{tabular}{
    lll
    S[table-format=1.2(1.2)]
    S[table-format=2.2(1.2)]
    S[table-format=1.2(1.2)]
    S[table-format=1.2(1.2)]
}
    \toprule
    Arch.  & Mod.   & Method & {$L_1 \downarrow$} & {$\nabla L_1 \downarrow$} & {PCC $\uparrow$} & {FD $\downarrow$} \\
    \midrule
    Mod.   & RMS    & Frame  &           1.82(0.01) &           4.03(0.24) &           0.42(0.01) & \bfseries 6.94(0.10) \\
    Synth  &        & LPF    & \bfseries 1.72(0.02) & \bfseries 2.91(0.01) & \bfseries 0.48(0.01) &           6.97(0.10) \\
           &        & Spline &           1.80(0.03) &           2.92(0.01) &           0.44(0.01) &           7.24(0.08) \\
    \cmidrule(lr){2-7}
           & SF     & Frame  &           2.67(0.06) &           7.85(0.38) & \bfseries 0.54(0.01) & \bfseries 3.66(0.04) \\
           &        & LPF    & \bfseries 2.63(0.04) & \bfseries 6.15(0.01) & \bfseries 0.54(0.01) &           3.75(0.03) \\
           &        & Spline &           2.97(0.08) &           6.20(0.01) &           0.46(0.01) &           3.92(0.03) \\
    \midrule
    Shan   & RMS    & Frame  &           1.85(0.02) &           4.06(0.18) &           0.41(0.01) & \bfseries 6.68(0.11) \\
    et al. &        & LPF    & \bfseries 1.75(0.02) & \bfseries 2.91(0.01) & \bfseries 0.45(0.01) &           6.92(0.07) \\
           &        & Spline &           1.81(0.03) &           2.93(0.01) &           0.43(0.01) &           7.19(0.06) \\
    \cmidrule(lr){2-7}
           & SF     & Frame  &           2.70(0.03) &           7.86(0.32) & \bfseries 0.53(0.01) & \bfseries 3.66(0.04) \\
           &        & LPF    & \bfseries 2.63(0.04) & \bfseries 6.15(0.01) &           0.52(0.01) &           3.76(0.03) \\
           &        & Spline &           3.02(0.07) &           6.21(0.01) &           0.45(0.01) &           3.95(0.02) \\
    \midrule
    Engel  & RMS    & Frame  &           2.18(0.02) &           4.31(0.14) &           0.22(0.01) &           7.54(0.09) \\
    et al. &        & LPF    & \bfseries 1.80(0.02) & \bfseries 2.95(0.01) & \bfseries 0.46(0.01) & \bfseries 7.24(0.08) \\
           &        & Spline &           1.86(0.02) &           2.96(0.01) &           0.43(0.01) &           7.36(0.07) \\
    \cmidrule(lr){2-7}
           & SF     & Frame  &           3.65(0.06) &           9.32(0.45) &           0.25(0.02) &           3.95(0.03) \\
           &        & LPF    & \bfseries 2.89(0.04) &           5.89(0.01) & \bfseries 0.43(0.01) & \bfseries 3.87(0.02) \\
           &        & Spline &           2.98(0.05) & \bfseries 5.88(0.01) &           0.39(0.01) &           3.92(0.02) \\
    \midrule
    All    & RMS    & Rand. S. &           2.20(0.01) &           2.96(0.01) &           0.22(0.01) &           7.82(0.02) \\
    \cmidrule(lr){2-7}
           & SF     & Rand. S. &           3.92(0.01) &           6.28(0.01) &           0.19(0.01) &           4.13(0.01) \\
    \bottomrule
\end{tabular}
\end{table}

\begin{table}[t]
\centering
\caption{Sound matching audio similarity metrics for real-world data.}
\label{tab:exp3_sound_matching}
\sisetup{
    reset-text-series = false, 
    text-series-to-math = true, 
    mode=text,
    tight-spacing=true,
    table-format=1.2,
    table-number-alignment=center,
    separate-uncertainty = true,
    detect-weight = true,
    detect-inline-weight = math,
}
\setlength{\tabcolsep}{4.2pt}
\begin{tabular}{
    ll
    S[table-format=2.2(1.2)]
    S[table-format=1.2(1.2)]
    S[table-format=3.2(2.2)]
    S[table-format=2.2(2.2)]
}
    \toprule
    Arch. & Method & {MSS $\downarrow$} & {MFCC $\downarrow$} & {FAD $\downarrow$} & {FAD $\downarrow$} \\
          &        &                    &                     & {MS-CLAP}          & {EnCodec}          \\
    \midrule
    Mod.   & Gran.     & \bfseries 1.47(0.01) & \bfseries 2.31(0.02) & \bfseries 311.24( 4.91) & \bfseries 29.21( 0.95) \\
    Synth  & Frame     &           1.60(0.05) &           2.56(0.10) &           361.52( 9.51) &           34.96( 2.64) \\
           & LPF       &           1.63(0.01) &           2.48(0.02) &           395.32( 9.72) &           37.81( 0.89) \\
           & Spline    &           1.67(0.01) &           2.58(0.04) &           411.73(17.39) &           40.19( 1.52) \\
           & Rand. S.  &           2.79(0.11) &           3.40(0.07) &           505.15(50.36) &           53.50(12.49) \\
    \midrule
    Shan   & Gran.     & \bfseries 1.48(0.01) & \bfseries 2.28(0.03) & \bfseries 306.53( 7.11) & \bfseries 28.48( 1.23) \\
    et al. & Frame     &           1.61(0.02) &           2.54(0.04) &           381.84(17.17) &           37.91( 2.08) \\
           & LPF       &           1.66(0.02) &           2.60(0.07) &           434.43(25.58) &           41.37( 1.77) \\
           & Spline    &           1.69(0.01) &           2.60(0.05) &           469.68(15.07) &           44.19( 1.12) \\
           & Rand. S.  &           3.06(0.24) &           3.53(0.11) &           509.76(43.82) &           83.02(17.31) \\
    \midrule
    Engel  & Gran.     & \bfseries 1.81(0.04) & \bfseries 2.99(0.05) & \bfseries 461.24(15.16) & \bfseries 35.04(1.11) \\
    et al. & Frame     &           2.06(0.05) &           3.27(0.04) &           667.88(27.87) &           57.69(1.94) \\
           & LPF       &           3.18(0.01) &           5.94(0.01) &           699.92( 9.75) &           91.29(0.89) \\
           & Spline    &           3.19(0.01) &           5.94(0.01) &           728.35(10.65) &           93.23(0.75) \\
           & Rand. S.  &           4.42(0.15) &           6.28(0.14) &           644.06(21.19) &           99.41(1.81) \\
    \bottomrule
\end{tabular}
\end{table}

\vspace{-2pt}
\section{Conclusion}
\label{sec:experiments}

We introduce a framework that uncovers modulations present in audio through extraction, parameterization, and DDSP synth sound matching.
We demonstrate its ability to discover unseen and approximate ground truth mod. signals in synthetic and real-world audio.
Our results show that a low-pass filter parameterization best balances expressive sound matching with human readability, while a spline parameterization maximizes interpretability.
We believe that creative applications with unconventional input audio and analyzing the symmetry of discovered modulations are compelling directions for future work.




\clearpage



\bibliographystyle{IEEEtran}
\bibliography{refs25}

\end{document}